\documentclass[doublecol,english,final]{epl2}
\usepackage[T1]{fontenc}
\usepackage[latin1]{inputenc}

\usepackage{ae} 
\usepackage{graphicx}
\usepackage{psfrag}
\usepackage{babel}
\usepackage{amssymb}
\usepackage{amsmath}

\title{Dynamic nuclear polarisation in biased quantum wires with spin-orbit interaction}

\author{V. Tripathi\inst{1} \and A. C. H. Cheung\inst{2} \and N. R. Cooper\inst{2}
\shortauthor{V. Tripathi \etal}}

\institute{
\inst{1}Department of Theoretical Physics, Tata Institute of Fundamental
Research, Homi Bhabha Road, Mumbai 400005, India \\
\inst{2}Theory of Condensed Matter Group, Cavendish Laboratory, Department
of Physics, University of Cambridge, J.~J.~Thomson Avenue, Cambridge CB3 0HE, United
Kingdom
}

\abstract{We propose a new method for dynamic nuclear polarisation in a quasi
one-dimensional quantum wire utilising the spin-orbit interaction, 
the hyperfine interaction, and a finite source-drain potential difference. 
In contrast with current methods, our scheme does not rely on external 
magnetic or optical sources which makes independent control of closely 
placed devices much more feasible.
Using this method, a significant polarisation of a few per cent is possible in 
currently available InAs wires which may be detected by conductance 
measurements. 
This may prove useful for nuclear magnetic resonance studies in 
nanoscale systems as well as in spin-based devices where external magnetic and
optical sources will not be suitable. 
}

\begin{document}
\maketitle

The ability to locally manipulate spin has been the subject of intense
research recently. In particular, local control of nuclear polarisation in
the vicinity of quantum dots and in wires has diverse uses such as in nuclear
magnetic resonance (NMR) studies \cite{wald} or for control of nuclear spins
in devices where external magnetic fields and/or optical sources will not be
suitable or convenient as for instance in non-magnetic spin-filtering
\cite{koga} and qubits \cite{kane}. In some very beautiful recent experiments
related to spin-based qubits, dynamically-generated nuclear polarisation has
been used to control the decoherence of electron spins in quantum dots
\cite{imamoglu,petta}. Nuclear polarisation can also be used to manipulate the
electron spin-splitting in quantum wires \cite{nesteroff} without using a
magnetic field. Local control of nuclear polarisation in quantum wells using a
gate contact has now been demonstrated \cite{sanada}; however, the
polarisation itself involved the use of an optical source. In this Letter, we
propose a new method for dynamically generating and controlling nuclear
polarisation, locally and \emph{without} using magnetic and optical sources, 
in a nonequilibrium quasi one-dimensional quantum wire with spin-orbit 
interaction. By local we mean the nuclear polarisation is within the wire and not 
in the bulk surroundings.
We argue that the role of spin-orbit interaction in creating a spin-polarised
electron distribution cannot in this scheme be replaced with an external
magnetic field. We estimate the magnitude and build-up time for the
polarisation as around $2.5\%$ and $200$s respectively in typical InAs quantum
wires which compares favourably with other methods. We discuss 
possible ways in which 
the nuclear polarisation may be detected and measured, as for instance 
through a measurement of the two-terminal 
conductance \cite{cooper07,nesteroff}. 

The Overhauser effect \cite{overhauser} is used for dynamically
polarising nuclei through their hyperfine coupling with electrons
using a magnetic field and an external radio source tuned to the electron
Zeeman splitting. 
The electron spin resonance (ESR) inducing radio
source tends to equalise (saturate) the numbers of spin up and down
electrons even though they have different Zeeman energies --- this
implies that the chemical potentials of the two spin species must
be different and the spin distribution is not in equilibrium. The
electron spins at the higher chemical potential ultimately relax with
a spin flip to equalise the chemical potentials; and some of these
spins can be exchanged with the nuclei. In this manner, a nonzero
nuclear polarisation is built up.
The key requirement for the Overhauser scheme is the creation of a 
nonequilibrium electron spin distribution. This usually involves pumping 
and saturation of
the electron spins. In the presence of a magnetic field, saturation
may be attained by, besides ESR, hot electrons \cite{feher}, or optically
through unpolarised light \cite{lampel}. Hot electrons have been
used to generate nuclear polarisation in InSb \cite{clark} subjected
to a Zeeman field, and more recently, in GaAs \cite{hoch}. In the
so-called optical Overhauser effect, electron spin pumping is achieved
using circularly polarised light \cite{lampel}. 
However these methods all involve
the use of physically large external sources of magnetic field and light making
local control of nuclear polarisation difficult, which is the main
motivation of our work.

Spin-orbit interaction offers one way where the electron spin degeneracy
can be lifted without using an external field. The lifting of electron
spin degeneracy by a spin-orbit interaction differs in many ways from
that in a magnetic field. In particular, no electron spin polarisation
$\sigma_{e}$ can be created solely through spin-orbit interaction
in a single sub-band quantum wire carrying a current \cite{zhai}.
However, it can be shown that in a quasi
one-dimensional model where the spin-orbit coupling mixes different
sub-bands, it is possible to obtain a finite $\sigma_{e}$ by passing
an electric current through the wire \cite{governale1}. This will
be the model we shall consider. Spin-orbit interaction in semiconductor
devices is usually of the Rashba or Dresselhaus kind. For simplicity,
our analysis focuses on a Rashba interaction, but our results will
hold even in the presence of a Dresselhaus interaction. 

For strong spin-orbit coupling, the left and right moving charges
in the lowest two sub-bands of transverse momentum quantisation can
become, at sufficiently large wave-vectors, completely spin-polarised
with opposite spin orientations for the two directions \cite{governale1}.
In the same multi-band model, we show that a small (partial) $\sigma_{e}$
occurs even when the spin-orbit coupling is weak, typical
in quantum wires. As the left and right-moving electrons have overall
opposite polarisations, an applied potential difference
creates a nonequilibrium electron spin distribution. Thereupon we
show how nuclear polarisation develops in the quantum wire due to
the hyperfine coupling of the nuclei with the nonequilibrium electrons. 

It is important to note that merely polarising the conduction electron 
spins \emph{does not} in general lead to dynamic nuclear polarisation. 
We discuss later that dynamic nuclear polarisation is not possible by simply 
using an external magnetic field to polarise the conduction electrons instead
of spin orbit interaction.

Let us first discuss how spin-orbit coupling in a quasi one-dimensional
wire can lead to electron spin polarisation. Consider a two-dimensional
gas (2DEG) of electrons in the $(x,z)$ plane. The geometry of the
quantum wire is such that the electrons are confined in the $y$ and
$z$ directions and the transport {}``channel'' is along the $x$
axis. We assume a hard wall confinement at $z=0$ and $z=W,$ and
at $y=0$ and $y=\delta.$ The Hamiltonian of the 2D electrons is
\begin{eqnarray}
H  =\frac{1}{2m}(p_{x}^{2}+p_{z}^{2})+V(z)+H_{SO},\text{ where}\label{eq:hamiltonian}\end{eqnarray}
\begin{eqnarray}
H_{SO}  =\frac{\hbar k_{SO}}{m}(\sigma_{z}p_{x}-\sigma_{x}p_{z})\label{eq:HSO}\end{eqnarray}
is the Rashba spin-orbit interaction arising from the 2DEG's asymmetric
confinement in the $y$-direction. Its interaction strength, $k_{SO}$,
may be controlled by an external gate voltage \cite{grundler}. $\sigma_{z}$
and $\sigma_{x}$ are Pauli matrices in the space of electron spin.

In the absence of the Rashba interaction, the scattering states are
labelled by the sub-band index $n$ of transverse momentum quantisation:
\begin{eqnarray}
\psi_{n\mathbf{k}\sigma}  =e^{ik_{x}x}\sin(k_{z}^{(n)}z)|\sigma\rangle,\label{eq:sc-states1}\end{eqnarray}
 where $k_{z}^{(n)}=n\pi/W,$ $n=1,\,2,\dotsc,$ and the energy eigenvalues
are $E_{n}^{(0)}(k_{x})=\hbar^{2}k_{x}^{2}/2m+(\hbar n\pi)^{2}/2mW^{2}.$
$|\sigma\rangle=|\uparrow\rangle,|\downarrow\rangle$ are the eigenstates
of $\sigma_{z}.$ The $\sigma_{z}p_{x}$ term in Eq.(\ref{eq:HSO})
respects this sub-band quantisation but the $\sigma_{x}p_{z}$ term
mixes sub-bands whose indices differ by an odd number as well as the
spin eigenstates of $\sigma_{z}p_{x}.$ It is convenient to regard
this mixing term as a perturbation, and the rest as the bare Hamiltonian.
Then, the spatial part of the bare eigenstates has the same form (Eq.(\ref{eq:sc-states1}))
as that in the absence of spin-orbit interaction. The matrix elements
of the second term in the spin-orbit interaction are 
\begin{equation}
H_{SO}^{n'\sigma',n\sigma}=\frac{i\hbar^{2}k_{SO}}{mW}\frac{2nn'}{n'^{2}-n^{2}}[1-(-1)^{|n'-n|}]\delta_{\sigma,-\sigma'}.
\end{equation}
The mixing is strongest for sub-bands whose indices differ by one;
and so, it is sufficient to consider the two lowest sub-bands \cite{mireles}.
In this approximation, the truncated Hamiltonian is expressible as
a $4\times4$ matrix in the basis $|n\sigma>=|1\uparrow\rangle,|2\downarrow\rangle,|2\uparrow\rangle,|1\downarrow\rangle$
(in that order); \begin{equation}
H_{\text{trunc.}}=\begin{pmatrix}E_{1\uparrow} & -i\Delta_{SO} & 0 & 0\\
i\Delta_{SO} & E_{2\downarrow} & 0 & 0\\
0 & 0 & E_{2\uparrow} & i\Delta_{SO}\\
0 & 0 & -i\Delta_{SO} & E_{1\downarrow}\end{pmatrix},\label{eq:H-2band}\end{equation}
 where \begin{eqnarray}
E_{n\uparrow(\downarrow)}  =\frac{\hbar^{2}k_{x}^{2}}{2m}+\frac{\hbar^{2}n^{2}\pi^{2}}{2mW^{2}}\pm\frac{\hbar^{2}k_{SO}k_{x}}{m},\label{eq:E-bare}\end{eqnarray}
 $n=1,2,$ and $\Delta_{SO}=\tfrac{8}{3}\tfrac{\hbar^{2}k_{SO}}{mW}.$
The eigenvalues of $H_{\text{trunc.}}$ are \begin{eqnarray}
\epsilon_{1a(2b)}  =\frac{E_{1\uparrow}+E_{2\downarrow}}{2}\mp\frac{1}{2}\sqrt{(E_{2\downarrow}-E_{1\uparrow})^{2}+4\Delta_{SO}^{2}},\label{eq:E1pm}\\
\epsilon_{1b(2a)}  =\frac{E_{1\downarrow}+E_{2\uparrow}}{2}\mp\frac{1}{2}\sqrt{(E_{2\uparrow}-E_{1\downarrow})^{2}+4\Delta_{SO}^{2}}.\label{eq:E2pm}\end{eqnarray}
The eigenvectors are also easily found.  In the spin
basis $|\uparrow\rangle,|\downarrow\rangle,$ we have \begin{eqnarray}
\psi_{1a}  =e^{ik_{x}x}\sqrt{\frac{2}{W}}N_{1a}\left(\begin{array}{c}
\sin(\pi z/W)\\
ig_{1a}(k_{x})\sin(2\pi z/W)\end{array}\right),\label{eq:psi-1u}\\
\psi_{1b}  =e^{ik_{x}x}\sqrt{\frac{2}{W}}N_{1b}\left(\begin{array}{c}
ig_{1b}(k_{x})\sin(2\pi z/W)\\
\sin(\pi z/W)\end{array}\right),\label{eq:psi-1d}\end{eqnarray}
 where the {}``mixing'' functions $g_{1a}$ and $g_{1b}$ are 
\begin{equation}
g_{1a}(k_{x})  =\frac{1}{2\Delta_{SO}}\left((E_{2\downarrow}-E_{1\uparrow})-\sqrt{(E_{2\downarrow}-E_{1\uparrow})^{2}+4\Delta_{SO}^{2}}\right)\label{eq:g1a}
\end{equation}
 and $g_{1b}(k_{x})=g_{1a}(-k_{x}),$ and the normalisations are $N_{1a(b)}=1/\sqrt{1+(g_{1a(b)}(k_{x}))^{2}}.$
Note that at sufficiently large \emph{negative} values of $k_{x},$
$\psi_{1a}\rightarrow\sqrt{2/W}e^{ik_{x}x}\sin(\pi z/W)|\uparrow\rangle.$
Likewise, at sufficiently large \emph{positive} values of $k_{x},$
we have $\psi_{1a}\rightarrow-i\sqrt{2/W}e^{ik_{x}x}\sin(2\pi z/W)|\downarrow\rangle.$
Similar scattering states have also been obtained for more general
Rashba interactions and a parabolic confining potential \cite{moroz}.

\begin{figure}
\psfrag{energy}{\(\text{Energy}/(\hbar^{2}/2mW^{2})\)}
\psfrag{1u}{\(1\uparrow\)}
\psfrag{1d}{\(1\downarrow\)}
\psfrag{2u}{\(2\uparrow\)}
\psfrag{2d}{\(2\downarrow\)}
\psfrag{kx}{\(k_x W\)}


\onefigure[width=6.0cm,keepaspectratio]{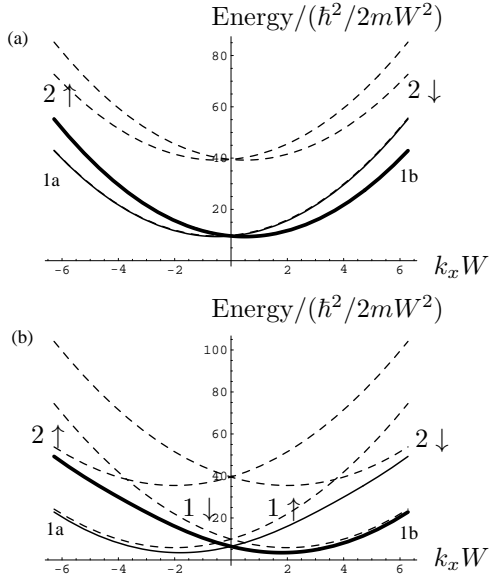}\psfragscanon

\caption{\label{cap:energylevel} Electron spin-polarisation $\sigma_{e}$
at large wave-vectors in quantum wires with Rashba interaction
--- dispersion curves in the two-band approximation for 
spin-orbit interaction strength $k_{SO}W= 0.5\, \text{(a) and }
k_{SO}W= 2\, \text{(b)}.$ Dotted curves: spin-orbit split bands in the
absence of sub-band mixing (Eq.(\ref{eq:E-bare})). Solid curves:
sub-band mixing lifts the degeneracy at $k_{x}\neq 0$ intersections.
(a) Weak coupling: $n=1$ and $n=2$ sub-bands intersection occurs at a
wave-vector much larger than that at the bottom of the $n=2$ sub-band.
Below this wave-vector, the lowest two sub-bands $1a$ and $1b$ are
oppositely polarised and $\sigma_{e}$ is small.  (b) Strong coupling:
the $n=1$ and $n=2$ sub-bands (dotted) intersect near the bottom of
the $n=2$ sub-band. When $k_{x} > 3\pi^{2}/(4k_{SO}W^{2})$ (dotted
intersection), electrons moving in a given direction in the lowest two
sub-bands $1a$ and $1b$ will have the same orientation of spin
polarisation.}
\end{figure}

Figure \ref{cap:energylevel} 
 explains how nearly complete electron spin-polarisation is possible in
quantum wires transmitting the lowest sub-bands 
under strong Rashba interaction.

We show that any Rashba interaction results in a finite
$\sigma_{e}.$ The electron spin polarisation
$\sigma_{e}=\langle\sigma_{z}(E)\rangle$ at a given energy $E$ for the $1a$
and $1b$ sub-bands is 
\begin{eqnarray}
  \sigma_e=\frac{1-(g_{1a}(E,k_{SO}))^{2}}{1+(g_{1a}(E,k_{SO}))^{2}}-\frac{1-(g_{1b}(E,k_{SO}))^{2}}{1+(g_{1b}(E,k_{SO}))^{2}}.\label{eq:spinpolar-gen}
\end{eqnarray}
The mixing functions $g_{1a,b}(k_{x})$ have been re-expressed as
functions of energy $E$ as $\sigma_{e}$ is to be calculated at a
fixed $E$ rather than a fixed $k_{x}.$ Clearly, in the absence of
inter sub-band mixing, $\langle\sigma_{z}(E)\rangle$ is zero. When
the spin-orbit interaction is weak ($k_{SO}W<1$), and the energy
is far from the region of sub-band mixing, the mixing functions $g_{1a(b)}$
will be small compared to unity and $\langle\sigma_{z}(E)\rangle\approx2[(g_{1b}(E,k_{SO}))^{2}-(g_{1a}(E,k_{SO}))^{2}].$
In this regime, the energy-wavevector relations for the $1a$ and
$1b$ sub-bands to $O(k_{SO}^{2})$ are respectively $k_{x}\approx-k_{SO}\pm\sqrt{2mE/\hbar^{2}-\pi^{2}/W^{2}}$
and $k_{x}\approx k_{SO}\pm\sqrt{2mE/\hbar^{2}-\pi^{2}/W^{2}},$ which
we can use to express $g_{1a}\approx-\Delta_{SO}/(E_{2\downarrow}-E_{1\uparrow})$
and $g_{1b}\approx-\Delta_{SO}/(E_{2\uparrow}-E_{1\downarrow})$ as
functions of $E:$\begin{eqnarray}
g_{1a}(E,k_{SO})=g_{1b}(E,-k_{SO})\approx-\frac{2mW^{2}\Delta_{SO}}{3\pi^{2}\hbar^{2}}\nonumber \\
\times\left[1+\frac{4k_{SO}W^{2}}{3\pi^{2}}\left(-k_{SO}\pm k_{F}^{1D}\right)\right].\label{eq:g1a-appr}\end{eqnarray}
 Here $k_{F}^{(1D)}=\sqrt{2mE/\hbar^{2}-\pi^{2}/W^{2}}$ is the 1D
Fermi wavevector in the $n=1$ sub-band in the absence of Rashba coupling.
It follows that for weak spin-orbit coupling, the electron spin polarisations
for right $(R)$ movers and left $(L)$ movers at energy $E$ are
respectively \begin{eqnarray}
\langle\sigma_{z}(E)\rangle_{R(L)}  \approx\mp6\left(\tfrac{16}{9\pi^{2}}\right)^{3}(k_{SO}W)^{3}(k_{F}^{(1D)}W).\label{eq:spinpolar-weak}\end{eqnarray}
 Note that unlike an external magnetic field, the right and left movers
have an \emph{opposite} net polarisation for both strong and weak
Rashba coupling. Thus, by imposing a net electrical current which
imbalances the left and right-moving electrons, we are able to satisfy
one of the requirements for the Overhauser effect; namely, we have
a non-equilibrium distribution of electron spin. Spin-orbit interaction
mediated electron spin polarisation can be controlled non-magnetically
also by applying a local strain \cite{kato}; however, this will not
fulfil our requirement of compact external sources. 

\begin{figure}
\psfrag{sigma}{$|\langle \sigma_{z} \rangle |$}
\psfrag{kso}{$k_{SO}W$}
\psfrag{E=10}{\small{$E=10$}}
\psfrag{E=20}{\small{$E=20$}}
\psfrag{E=30}{\small{$E=30$}}
\psfrag{E=40}{\small{$E=40$}}

\onefigure[width=8.0cm,keepaspectratio]{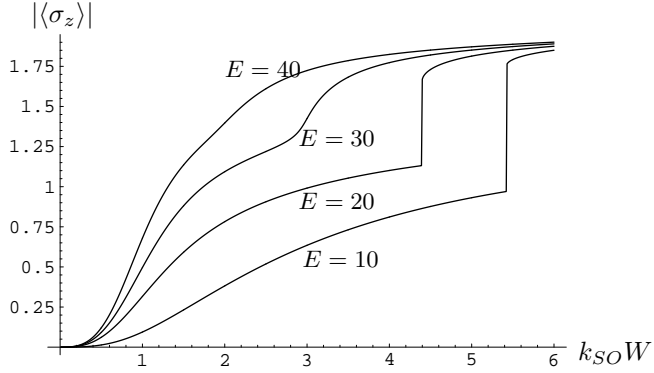}\psfragscanon

\caption{\label{cap:electronpolar} The electron spin polarisation (given by Eq.(\ref{eq:spinpolar-gen})) for the lowest two sub-bands for different values of the energy $E$ (measured in units of $\hbar^2/2mW^2$), and given chirality, plotted as a function of the dimensionless spin-orbit interaction strength $k_{SO}W$. The fraction of nuclei polarised at saturation, $\langle I_z\rangle/I,$ is of the order of $I |\langle \sigma_z \rangle |$. Actual InAs/GaSb quantum wires are typically associated with spin-orbit interaction strengths in the region $k_{SO}W\lesssim 1$.}
\end{figure}

In Figure \ref{cap:electronpolar} we show a numerical calculation for the electron spin polarisation $|\langle \sigma_{z} \rangle |$ for the lowest two sub-bands ($1a$ and $1b$) as a function of the dimensionless spin-orbit interaction $k_{SO}W$ at different values of the energy $E$, using Eq.(\ref{eq:spinpolar-gen}). The range of values of energy shown spans, approximately, the range of Fermi energy over which only the two lowest sub-bands ($1a$ and $1b$) are occupied. Note that the polarisation undergoes a significant change when $k_{SO}W$ reaches some large value; this is especially true at low energies. One can understand this change by looking at the spin structure of the $1a$ and $1b$ sub-bands in Figure \ref{cap:energylevel}b. When the energy is small enough, the two sub-bands have opposite spin polarisations, and total polarisation is the difference of the polarisations of the two. As $k_{SO}W$ is increased, the sub-bands move down in energy until, eventually, the energy will exceed the value at which both sub-bands have the same polarisation, which causes a significant increase in $|\langle \sigma_{z} \rangle |$.

Now we show how a nonzero nuclear polarisation can be created with
a finite source-drain potential difference but without using an external
magnetic field. Consider the hyperfine interaction of the electron
spin density $\mathbf{s}(\mathbf{r}_{i})$ with the host nuclear moments
$\mathbf{I}_{i}$ at the centre of the quantum wire,\begin{eqnarray}
H_{\text{hyp}}  =A\sum_{i}[2(s^{+}(\mathbf{r}_{i})I_{i}^{-}+s^{-}(\mathbf{r}_{i})I_{i}^{+})+s^{z}(\mathbf{r}_{i})I_{i}^{z}],\label{eq:hyperfine}\end{eqnarray}
where $A\equiv(4/W\delta)A_{3d},$ where $A_{3d}$ is the bulk hyperfine
constant, $4/W\delta$ is the electron density at the centre of the
wire, and $s^{+}=(s_{x}+is_{y})/2,$ etc. When the Rashba interaction
is strong and the Fermi energy is large enough, right movers at the
Fermi surface are nearly completely polarised spin down and the corresponding
left movers are polarised spin up. When a right-moving electron at
the Fermi surface, say in the $1b$ state (i.e.~spin down), exchanges
its spin with a nucleus because of the hyperfine coupling, the final
state of the electron can either be a higher energy sub-band such
as $2b$ (not shown in Figure \ref{cap:energylevel}) with $k_{x}$
in the same direction, or a sub-band such as $1a$ with $k_{x}$ in
the opposite direction with a similar energy. The first process cannot
occur with energy conservation, unless there is simultaneous energy
absorption from the heat bath; hence it is small (exponentially activated)
at low temperature. The latter process involves a backscattering which
can occur while scattering from the nucleus. If the chemical potential
$\mu_{L}$ of the left contact is higher than the chemical potential
$\mu_{R}$ of the right contact, then after backscattering with a
spin flip, the electron can have a higher energy than the chemical
potential of the right contact. Spin-flip backscattering of the left-moving
electrons is suppressed because $\mu_{R}<\mu_{L}.$ We thus have a
situation where electrons exchanging spins with the nuclei scatter
preferentially from a spin down state.

The rate of change of transition probability for a hyperfine mediated
scattering, say with $\Delta m_{I}=-1,$ is
\begin{eqnarray}
w_{kk'}  =\frac{8\pi}{\hbar}GA^{2}\nu_{f},
\label{eq:fermigolden1}
\end{eqnarray}
where $\nu_{f}=m/(2\pi\hbar^{2}k_{F}^{(1D)})$ is the (1D) density
of final states and $G=\tfrac{1}{2I+1}\sum_{m=-I}^{I}|I_{m-1,m}^{-}|^{2}=\tfrac{2}{3}I(I+1),$
is the appropriate average value of $|I^{-}|^{2}$ assuming that the
nuclear moments are unpolarised. We now obtain an expression for the
rate of electron flips $w(R\downarrow;L\uparrow)$ from the right-moving
$|\downarrow\rangle$ state to the left-moving $|\uparrow\rangle$
state following the general prescription of Ref. \cite{overhauser2}.
The distribution functions of the right-moving and left-moving electrons
are $f_{R(L)}^{n,s}(k_{x})=\frac{1}{e^{(\epsilon_{n,s}(k_{x})-\mu_{L(R)})/k_{B}T}+1}.$
The energies $\epsilon_{n,s}(k_{x})$ have been obtained in Eq.(\ref{eq:E1pm})
and Eq.(\ref{eq:E2pm}). 
We have not considered magnetic fields here
but that can also be incorporated in principle. 
The rate of electron
flips per nucleus from $R\downarrow$ to $L\uparrow$ is 
\begin{equation}
w(R\downarrow;L\uparrow)  =\sum_{n,s}\int_{k_{\text{min}}}^{k_{\text{max}}}\frac{dk_{x}}{2\pi}w_{kk'}f_{R}^{n,s}(k_{x})[1-f_{L}^{n',s'}(k'_{x})].\label{eq:tr-prob1}
\end{equation}
Here $k_{\text{min}}$ and $k_{\text{max}}$ correspond to the minimum
and maximum values of $k_{x}>0$ where the electrons can be considered
to be nearly completely polarised spin-down. $k_{\text{min}}$ is
approximately given by the value of $k_{x}$ at which the $1\uparrow$and
$2\downarrow$ bands cross (see Figure \ref{cap:energylevel}). We
assume that the temperature is low compared to the inter sub-band
separation as well as the potential difference and choose the left
and right chemical potentials to lie in the energy band where spin
polarisation is possible. The energy conservation condition for the
above transition is $\epsilon_{n,s}(k_{x})=\epsilon_{n',s'}(k'_{x}).$
Starting with unpolarised nuclei, the \emph{initial} rate of change
of the polarisation $\langle I_{z}\rangle$ per nucleus at the centre
of the wire is $-[w(R\downarrow;L\uparrow)-w(L\uparrow;R\downarrow)]$, i.e.
\begin{eqnarray}
 \frac{d\langle I_{z}\rangle}{dt}  =-\frac{8\pi}{\hbar}GA^{2}(\mu_{L}-\mu_{R})\sum_{ns,n's'}\nu_{ns}\nu_{n's'}.\label{eq:nuclearpol}
\end{eqnarray}
Here scattering from $L\uparrow$ to $R\downarrow$ is suppressed
because $\mu_{R}<\mu_{L}.$ For $I=1/2$ or at long times in general,
$\langle I_{z}\rangle+I$ will decay exponentially. A spin-down nuclear
polarisation is thus built up.

When spin-orbit coupling is weak, and the Fermi energy is far from
the (avoided) intersection of the sub-bands, we have shown earlier
that the right movers in the $n=1$ sub-bands can have either spin
orientation with a small excess of spin down electrons. Likewise the
left movers have a small excess of spin up electrons. Thus spin-flip
backscattering of spin-down right movers is not completely cancelled
by the spin-flip backscattering of the spin-up right movers. Forward
scattering does not contribute as in this case the two competing spin
flips occur at the same rate. The initial rate of change of $\langle I_{z}\rangle,$
therefore, is suppressed by a factor $|\langle\sigma_{z}(\epsilon_{F})\rangle|.$
The initial rate $T_{n}^{-1}$ of build-up of the polarisation $\langle I_{z}\rangle$
of a nucleus is \begin{eqnarray}
T_{n}^{-1}  =\frac{8\pi}{\hbar}GA^{2}|\langle\sigma_{z}(\epsilon_{F})\rangle|\Delta\mu\!\!\!\sum_{ns,n's'}\!\!\!\nu_{ns}\nu_{n's'}\nonumber \\
  \approx\frac{3GA_{3d}^{2}}{4}\frac{16^{5}}{9^{3}\pi^{7}}\frac{m^{2}k_{SO}^{3}}{\hbar^{5}k_{F}^{(1D)}}\left(\frac{W}{\delta}\right)^{2}\Delta\mu,\label{eq:Tn}\end{eqnarray}
where $\Delta\mu=|\mu_{L}-\mu_{R}|$ and $\nu_{n,s}\approx m/(2\pi\hbar^{2}k_{F}^{(1D)})$.
We used Eq.(\ref{eq:spinpolar-weak}) for $|\langle\sigma_{z}(\epsilon_{F})\rangle|\ll1,$
which is the expected experimental situation. We considered only the
lowest sub-bands. At finite temperatures exceeding $|\Delta\mu|/k_{B},$
the polarisation build-up rate decreases by an amount proportional
to the temperature owing to Korringa relaxation. Eq.(\ref{eq:Tn})
is the main result of the paper. 

\begin{figure}
\psfrag{Tn}{$T_{n}\,\rm{(s)}$}
\psfrag{kso}{$k_{SO}W$}
\psfrag{E=10}{\small{$E=10$}}
\psfrag{E=40}{\small{$E=40$}}
\psfrag{logTn}{$\log(T_{n})$}
\psfrag{logkso}{$\log(k_{SO}W)$}

\onefigure[width=8.0cm,keepaspectratio]{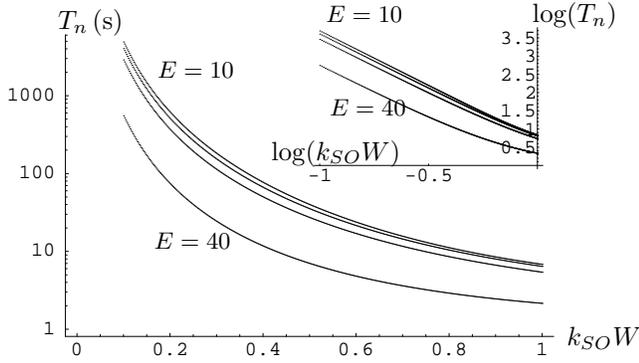}\psfragscanon

\caption{\label{cap:builduptime} The initial build up time (measured in seconds) for nuclear polarisation, $T_n$, given by the first of Eqs.(\ref{eq:Tn}), is shown here as a function of $k_{SO}W$ for dimensionless energies $E=10,\,20,\,30,$ and $40$, nuclear spin $I=9/2$, and a potential difference of $1\rm{mV}$ between the left and right leads. The inset plot of 
$\log(T_n)$ versus $\log(k_{SO}W)$ clearly reveals the $k_{SO}^3$ dependence of the nuclear polarisation for $k_{SO}W < 1$ obtained analytically in Eq.(\ref{eq:Tn}). The example discussed at the end of the paper approximately corresponds to $k_{SO}W=0.3$.}
\end{figure}

Figure \ref{cap:builduptime} shows a numerical calculation for $T_{n}$ for a InAs/GaSb quantum wire as a function of $k_{SO}W$ for four different values of the energy at a fixed potential difference of $1\rm{mV}$ and $\delta = 5\rm{nm}$.  The log-log plot in the inset shows that the $k_{SO}^3$ law obtained in Eq.(\ref{eq:Tn}) is indeed obeyed.

Complete nuclear polarisation is not possible in our scheme when $|\langle\sigma_{z}(\epsilon_{F})\rangle|<1$ per sub band.
Consider the worst case scenario, $I=1/2.$ Nuclear polarisation stops
increasing once $w(R\downarrow;L\uparrow)p_{\uparrow}-w(R\uparrow;L\downarrow)p_{\downarrow}=0,$
where $p_{\uparrow(\downarrow)}$ is the probability that the nucleus
is spin up (down). It is easy to see that this amounts to a steady
state nuclear polarisation $\langle I_{z}\rangle\sim|\langle\sigma_{z}(\epsilon_{F})\rangle|.$
However, for larger values of $I,$ a nuclear transition $m_{I}\rightarrow(m_{I}-1)$
does not, except in the lowest two states, cause a corresponding increase
in the rate of the undesirable $(m_{I}-1)\rightarrow m_{I}$ transition.
Larger polarisations are thus possible for $I>1/2:$ $\langle I_{z}\rangle/I\sim I|\langle\sigma_{z}(\epsilon_{F})\rangle|$
for small $|\langle\sigma_{z}(\epsilon_{F})\rangle|.$ Including more
sub-bands will further enhance both $T_{n}^{-1}$ and $|\langle\sigma_{z}(\epsilon_{F})\rangle|.$

The role of Rashba coupling in the build-up of nuclear polarisation
cannot be substituted by an external (Zeeman) magnetic field. 
In that case, under conditions of a non-zero (but small) bias, one can show
that within linear response, the probability of a nuclear spin-flip
process involving scattering of a spin-up right-mover into a spin-down
left mover is identical to the probability of a nuclear spin-flip
process involving scattering of a spin-down right-mover into a spin-up
left mover. These two processes compensate leading to no net build-up
of nuclear polarisation.

The proposed method works even in the presence of a Dresselhaus spin-orbit
interaction, say, of the form $H_{D}=\frac{\hbar k_{D}}{m}(p_{z}\sigma_{z}-p_{x}\sigma_{x}).$
This is most easily seen by going over to a basis consisting of eigenstates
of $\sigma_{x}.$ Then the Pauli matrices transform as $(\sigma_{x},\sigma_{y},\sigma_{z})\rightarrow(\sigma_{z},-\sigma_{y},\sigma_{x}),$
and the Dresselhaus interaction changes to a Rashba interaction and
our earlier treatment can be repeated. Hence if only Dresselhaus is
present, the spins will align along the $x$ direction. If both Rashba
and Dresselhaus are present, then the spin polarisation will be aligned
along some direction in the $x$--$z$ plane. If the Dresselhaus interaction
also contains a $\sigma_{y}p_{y}$ term, that will cause a mixing
of sub-bands in a direction perpendicular to the 2DEG. Since the confinement
in this direction is stronger than the confinement in the plane, the
separation of the sub-bands will be large and their mixing will be
small compared to the in-plane sub-bands.

Consider a typical InAs/GaSb heterostructure with effective electron
mass $m=0.027m_{e},$ $I_{\text{As}}=3/2$ and $I_{\text{In}}=9/2,$
and confinement ratio $(W/\delta)^{2}=10$ (typical ratio of sub-band
energy spacings in GaAs devices). The atomic density is $3.6\times10^{28}\:\mathrm{m}^{-3},$
hyperfine couplings are of the order of $100\:\mu{\mathrm{eV}}$ per
atom (see for e.g.\ Ref.\
\cite{braun}), and the Rashba coupling from Ref.\
\cite{grundler} is $k_{SO}=1.4\times10^{7}\:\mathrm{m}^{-1}.$
The average value of $G$ is $9.5.$ For a small potential difference
$\Delta\mu=1\:{\mathrm{meV}}\lesssim\epsilon_{F},$ and using $\epsilon_{F}=\hbar^{2}(k_{F}^{(1D)})^{2}/2m\sim\Delta\mu,$
we estimate from Eq.(\ref{eq:Tn}) $T_{n}\approx200{\rm s}.$ Estimating 
$W\sim 1/k_{F}^{(1D)}$, we have $\langle I_{z}\rangle /I \approx 2.5\times 10^{-2}$ for In atoms. The
build-up rate is very sensitive to the spin-orbit coupling strength
and the confinement asymmetry. The small number of polarised spins may pose a
considerable problem for detection by conventional NMR. 
Though the situation can be further improved by having multiple quantum wires 
parallel to each other, this is not ideal.  

We would like to suggest two methods of detection.  Firstly, because $\langle I_{z}\rangle \ne 0$, the
final term of the hyperfine Hamiltonian \eqref{eq:hyperfine} produces an
effective magnetic field.  By applying an in-plane magnetic field as a
reference, it is possible to infer the strength of the effective hyperfine 
field and hence the polarisation from two-terminal conductance measurements 
\cite{nesteroff,cooper07}.Secondly, since
polarised radioactive nuclei have an anisotropic decay pattern
\cite{bloembergen}, it should be possible to detect polarisation by
measuring the beta decay statistics \cite{winnacker}.  In this case,
$^{114}$In possesses a half-life of 50 days,
and may prove suitable.  

Dipolar interactions of the nuclei will reduce the steady state
nuclear polarisation through spin diffusion out of the boundary of the
quantum wire, and spin-flip processes that do not conserve total
spin. Small magnetic fields can be used to strongly suppress such
processes\cite{maletinsky}.

In conclusion, we have shown how a gate-controlled local dynamic nuclear
polarisation is possible in a quasi one-dimensional wire with spin-orbit
interaction subjected to a finite source-drain potential. A significant 
polarisation of a few per cent can be easily attained in present InAs 
devices in about $200$s, and may be detected by conductance measurements 
\cite{nesteroff,cooper07}. 
The polarisation is very sensitive to the strength of (gate-controlled) 
spin-orbit coupling as well as the asymmetry of confinement in the 
quantum wire.  This method may prove useful in NMR studies in semiconductor
devices and in the local control of electron spin decoherence in spin-based 
devices.
 
\acknowledgments

VT thanks the support of TIFR and a DST Ramanujan Fellowship (sanction
no.\ 100/IFD/154/2007-08). ACHC thanks Trinity college, Cambridge for
support. NRC acknowledges support by EPSRC grant GR/S61263/01.


\begin{thebibliography}{10}
\bibitem{wald}Wald K. R., Kouwenhoven L. P., McEuen P. L., van der Vaart N. C. 
        and Foxon C. T., \emph{Phys. Rev. Lett.} \textbf{73} (1994) 1011.
\bibitem{koga}Koga T., Nitta J., Takayanagi H. and Datta S., 
        \emph{Phys. Rev. Lett.} \textbf{88} (2002) 126601.
\bibitem{kane}Kane B. E., \emph{Nature} \textbf{393} (1998) 133.
\bibitem{imamoglu}Imamoglu A., Knill E., Tian L. and Zoller P., 
        \emph{Phys. Rev. Lett.} \textbf{91} (2003) 017402.
\bibitem{petta}Petta J. R., Taylor J. M., Johnson A. C., Yacoby A., 
        Lukin M. D., Marcus C. M., Hanson M. P. and Gossard A. C., 
        arXiv:0709.0920 [cond-mat.mes-hall] preprint, 2007. 
\bibitem{nesteroff}Nesteroff J. A., Pershin Y. V. and Privman V., 
        \emph{Phys. Rev. Lett.} \textbf{93} (2004) 126601. 
\bibitem{sanada}Sanada H., Matsuzaka S., Morita K., Hu C. Y., Ohno Y. 
        and Ohno H., \emph{Phys. Rev. Lett.} \textbf{94} (2005) 097601. 
\bibitem{cooper07}Cooper N. R. and Tripathi V., \texttt{arXiv:0710.4302} (2007). 
\bibitem{overhauser}Overhauser A. W., \emph{Phys. Rev.} 
        \textbf{92} (1953) 411. 
\bibitem{feher}Feher G., \emph{Phys. Rev. Lett.} \textbf{3} (1959) 135. 
\bibitem{lampel}Lampel G., \emph{Phys. Rev. Lett.} \textbf{20} (1968) 491. 
\bibitem{clark}Clark W. G. and Feher G., \emph{Phys. Rev. Lett.} 
        \textbf{10} (1963) 134. 
\bibitem{hoch}Hoch M. J. R., Lu J., Kuhns P. L., Moulton W. G. and 
        Reyes A. P., \emph{Phys. Rev. B} \textbf{72} (2005) 233204. 
\bibitem{zhai}Zhai F. and Xu H. Q., \emph{Phys. Rev. Lett} 
        \textbf{94} (2005) 246601. 
\bibitem{governale1}Governale M. and Z\"{u}licke U., \emph{Phys. Rev. B} 
        \textbf{66} (2002) 073311; Eto M., Hayashi T. and Kurotani Y., 
        \emph{J. Phys. Soc. Jpn.} \textbf{74} (2005) 1934; 
        Perroni C. A., Bercioux D., Ramaglia V. M. and Cataudella V., 
        \emph{J. Phys.: Condens. Matter} \textbf{19} (2007) 186227.  
\bibitem{grundler}Grundler D., \emph{Phys. Rev. Lett.} 
        \textbf{84} (2000) 6074; 
        Nitta J., Akazaki T., Takayanagi H. and Enoki T., 
        \emph{Phys. Rev. Lett.} \textbf{78} (1997) 1335. 
\bibitem{mireles}Mireles F. and Kirczenow G., \emph{Phys. Rev. B} 
        \textbf{64} (2001) 024426. 
\bibitem{moroz}Moroz A. V. and Barnes C. H. W., \emph{Phys. Rev. B} 
        \textbf{60} (1999) 14272. 
\bibitem{kato}Kato Y. K., Myers R. C., Gossard A. C. and Awschalom D. D., 
        \emph{Phys. Rev. Lett.} \textbf{93} (2004) 176601.
\bibitem{overhauser2}Overhauser A. W., \emph{Phys. Rev.} 
        \textbf{89} (1953) 689. 
\bibitem{braun}Braun P.-F., Marie X., Lombez L., Urbaszek B., Amand T., 
        Renucci P., Kalevich V. K., Kavokin K. V., Krebs O., Voisin P. 
        and Masumoto Y., \emph{Phys. Rev. Lett.} \textbf{94} (2005) 116601.
\bibitem{bloembergen}Bloembergen N. and Temmer G. M., \emph{Phys. Rev.} 
        \textbf{89} (1953) 883.
\bibitem{winnacker}Winnacker A., Ackermann H, Dubbers D, Mertens J. and 
        von Blanckenhagen P., \emph{Z. Physik} 
        \textbf{244} (1971) 289. 
\bibitem{maletinsky}Maletinsky P., Badolato A. and Imamoglu A., 
        \emph{Phys. Rev. Lett.} \textbf{99} (2007) 056804.
\end{thebibliography}
\end{document}